\documentclass[12pt]{article}

\usepackage{amsmath}
\usepackage{graphicx}
\usepackage{booktabs}
\usepackage{amsfonts}
\usepackage{hyperref}
\usepackage{amssymb}

\renewcommand{\theequation}
{\arabic{section}.\arabic{equation}}

\makeatletter
\def\eqnarray{ \stepcounter{equation} \let\@currentlabel=\theequation
 \global\@eqnswtrue
 \global\@eqcnt\z@
 \tabskip\@centering
 \let\\=\@eqncr
 $$\halign to \displaywidth\bgroup\@eqnsel\hskip\@centering
 $\displaystyle\tabskip\z@{##}$&\global\@eqcnt\@ne
 \hfil$\displaystyle{{}##{}}$\hfil
 &\global\@eqcnt\tw@$\displaystyle\tabskip\z@{##}$\hfil
 \tabskip\@centering&\llap{##}\tabskip\z@\cr}
\makeatother

\makeatletter
\def\@arrayacol{\edef\@preamble{\@preamble \hskip .5\arraycolsep}}
\def\array{\let\@acol\@arrayacol \let\@classz\@arrayclassz
\let\@classiv\@arrayclassiv \let\\\@arraycr\def\@halignto{}\@tabarray}
\makeatother

\makeatletter
\newcounter{subeqncnt}
\def\thesubeqncnt{\alph{subeqncnt}}
\def\subequations{\begingroup%
   \stepcounter{equation}\edef\@tempa{\theequation}%
   \let\c@equation\c@subeqncnt\c@subeqncnt\z@
   \edef\theequation{\@tempa\noexpand\thesubeqncnt}}

\makeatother

\newcommand{\be}{\begin{equation}}
\newcommand{\ee}{\end{equation}}

\newcommand{\bea}{\begin{eqnarray}}
\newcommand{\eea}{\end{eqnarray}}
\newcommand{\nn}{\nonumber}

\def\CM {{\cal M}}

\def\CR {{\cal R}}

\begin{document}

\setlength{\baselineskip}{7mm}
\begin{titlepage}
 \begin{flushright}
{\tt NRCPS-HE-24-2015}
\end{flushright}

\begin{center}
{\Large ~\\{\it   Anosov C-systems\\ and \\Random Number Generators
\vspace{1cm}

}

}

\vspace{1cm}

{\sl George Savvidy

\bigskip
\centerline{${}$ \sl Institute of Nuclear and Particle Physics}
\centerline{${}$ \sl Demokritos National Research Center, Ag. Paraskevi,  Athens, Greece}
\bigskip

}
\end{center}
\vspace{30pt}

\centerline{{\bf Abstract}}
We are developing further our earlier suggestion to use hyperbolic Anosov C-systems
for the Monte-Carlo simulations in high energy particle physics. The hyperbolic dynamical systems 
have homogeneous instability of all trajectories and as such they have 
mixing of all orders, countable Lebesgue  spectrum and  positive Kolmogorov entropy.
These extraordinary ergodic properties follow from the C-condition introduced by Anosov.
The C-condition defines a rich class of dynamical systems which span an open set in 
the space of all dynamical systems.  The important property of 
C-systems is that they have a countable  set of everywhere dense periodic trajectories
and that their density exponentially increases with entropy. 
Of special interest are C-systems 
that are  defined on a high dimensional  torus. The C-systems on a torus are 
perfect candidates to be used for Monte-Carlo simulations. 
Recently  an efficient algorithm was found, which  allows very fast generation of long 
trajectories of the C-systems. These trajectories have high quality statistical properties 
and we are suggesting to use them for the QCD lattice simulations and at high energy particle physics.

\vspace{12pt}

\centerline{ {\Large \it  Dedicated to the memory of Professor Dmitri   Anosov }}

\noindent

\end{titlepage}




\pagestyle{plain}

\section{\it Introduction}

\begin{figure}
\centering
\includegraphics[width=7cm]{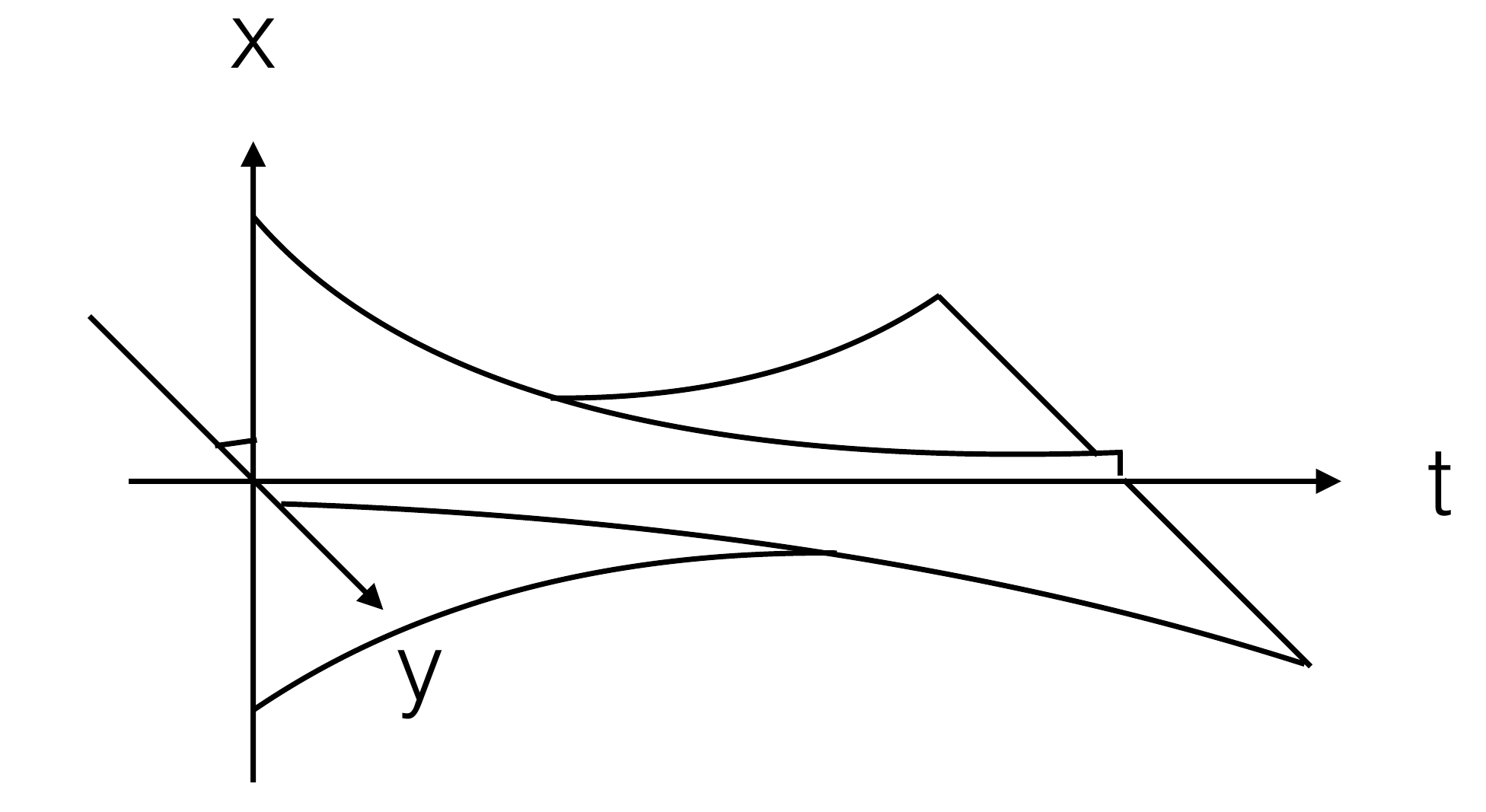}
\centering
\caption{The integral curves in the case of saddle point 
$\dot{x} =-x,~~~\dot{y}=y.$ The t axis is the intersection of two planes 
of integral curves which  are approaching it at $t \rightarrow +\infty$, the plane (x,t)
and at $t \rightarrow -\infty$, the plane (y,t). The rest of the integral curves 
are expanding at $t \rightarrow +\infty$, as well as at $t \rightarrow -\infty$.
A similar behaviour takes place in the neighbourhood of almost all  trajectories 
of the C-system \cite{anosov}. } 
\label{fig1}
\end{figure}

In the fundamental work on geodesic flows on closed
Riemannian manifolds $V^n$ of negative curvature \cite{anosov} Dmitri Anosov
pointed out  that the basic property of the geodesic flow on such manifolds 
is a {\it uniform  instability of all trajectories}, 
which in physical terms means that {\it in  the 
neighbourhood  of every fixed trajectory the trajectories 
behave similarly to the trajectories in the neighbourhood of a saddle point} (see Fig. \ref{fig1}).
In other words, the hyperbolic instability of the 
dynamical system  $\{ T^t \} $ which is defined by the equations \footnote{It is 
understood that the phase space manifold   $W^m$ is equipped by the invariant   
Liouville  measure \cite{anosov}.} 
\be\label{hyperbolic}
\dot{w} = f(w)
\ee
takes place for all solutions $\delta w \equiv \omega$ of the deviation equation  
\be
 \dot{\omega} = {\partial f \over \partial w } \bigg\vert_{w(t)=T^t w} \omega
\ee
in the neighbourhood of each phase trajectory  $w(t)=T^t w$, where $w\in W^m$.

The exponential instability of geodesics on Riemannian manifolds of 
constant negative curvature 
has been studied by many authors, beginning with Lobachevsky and Hadamard 
and especially  by  Hedlund and Hopf \cite{hedlund,hopf}.
The concept of exponential instability of {\it every trajectory} of a
dynamical system appears to be extremely rich and Anosov suggested 
to elevate it into a fundamental property of a new class of dynamical systems
which he called C-systems\footnote{The letter C is used because these systems 
fulfil the "C condition" (\ref{ccondition})\cite{anosov}. }. 
The brilliant idea to consider dynamical systems 
which have  {\it local and homogeneous  hyperbolic instability of all trajectories }
is appealing to the intuition and  has  very deep  physical content.   
The richness of the concept 
is expressed by the fact  that  the C-systems 
occupy a nonzero volume  in the space of dynamical 
systems \cite{anosov}\footnote{This is in a contrast 
with the integrable systems where  only a part of the 
trajectories remain on the invariant tori in accordance with the  KAM 
theorem.}.  

Anosov provided an extended list  of   C-systems \cite{anosov}. 
Important examples of the C-systems are: i)   
the geodesic flow on the Riemannian manifolds of {\it variable negative curvature} and 
ii) C-cascades - the iterations of the hyperbolic automorphisms of tori.

Already from the intuitive definition of the C-systems 
it is clear  that the  C-systems have very strong instability of 
their trajectories and in fact the instability is as strong as it can be in principle \cite{anosov1}.
The distance between infinitesimally close trajectories increases 
exponentially and on a compact phase space of the dynamical system 
this leads to the uniform distribution of almost all trajectories
over the whole phase space.  Thus
the dynamical systems which fulfil the C-condition have very extended and 
rich  ergodic properties \cite{anosov}.  As such they have 
{\it mixing of all orders, countable Lebesgue  spectrum and   positive entropy} \cite{anosov}.

These properties follow from the theorems proven by Anosov  and
stating that the  C-systems, with one exception\footnote{These are the  C-flows 
produced  by some standard construction from the corresponding C-cascades \cite{anosov} (see also Appendix).},  
are in fact also Kolmogorov K-systems \cite{kolmo,kolmo1,sinai3} .  
Thus the C-condition, in most of the  cases, is a sufficient condition for the dynamical 
system to be a K-system as well.
In this sense the C-systems provide extended and rich list of concrete examples of K-systems
\cite{anosov}. At the same time there are examples of  K-systems which are not C-systems, 
therefore in the  space of vector fields $\{f(w)\}$ defining 
the dynamical systems $\dot{w}= f(w)$, 
these two classes of dynamical systems are overlapping 
but are not identical  (see also Footnote (\ref{note})). 

The other important property of the C-systems 
is that in "between" the uniformly distributed trajectories 
there is  {\it a countable set of periodic trajectories.   The set of points 
on the periodic trajectories is everywhere dense in the phase space manifold $W^{m}$}
\cite{anosov}. 
The periodic trajectories and uniformly distributed trajectories 
are filling out the phase space of a C-system in a way very similar to 
the rational and irrational numbers on the real line.

Let us define the C-condition for the dynamical systems with discrete time \cite{anosov}.
A {\it cascade}  on the m-dimensional compact phase space $W^{m}$ is induced by 
the diffeomorphisms $T: W^m \rightarrow W^m$.  The iterations
are defined by  a repeated action of the operator  
$\{ T^n, -\infty < n < +\infty  \}$, where $n$ is an  integer number.  
The tangent space at 
the point $w \in W^m$ is  denoted by $R^m_{w}$ and the 
tangent vector bundle by $\CR(W^m)$. 
The diffeomorphism $\{T^n\}$ induces the mapping
of the tangent spaces $\tilde{T}^{n}: R^m_w \rightarrow R^m_{ T^{n}w}$.
\begin{figure}
  \centering
\includegraphics[width=8cm]{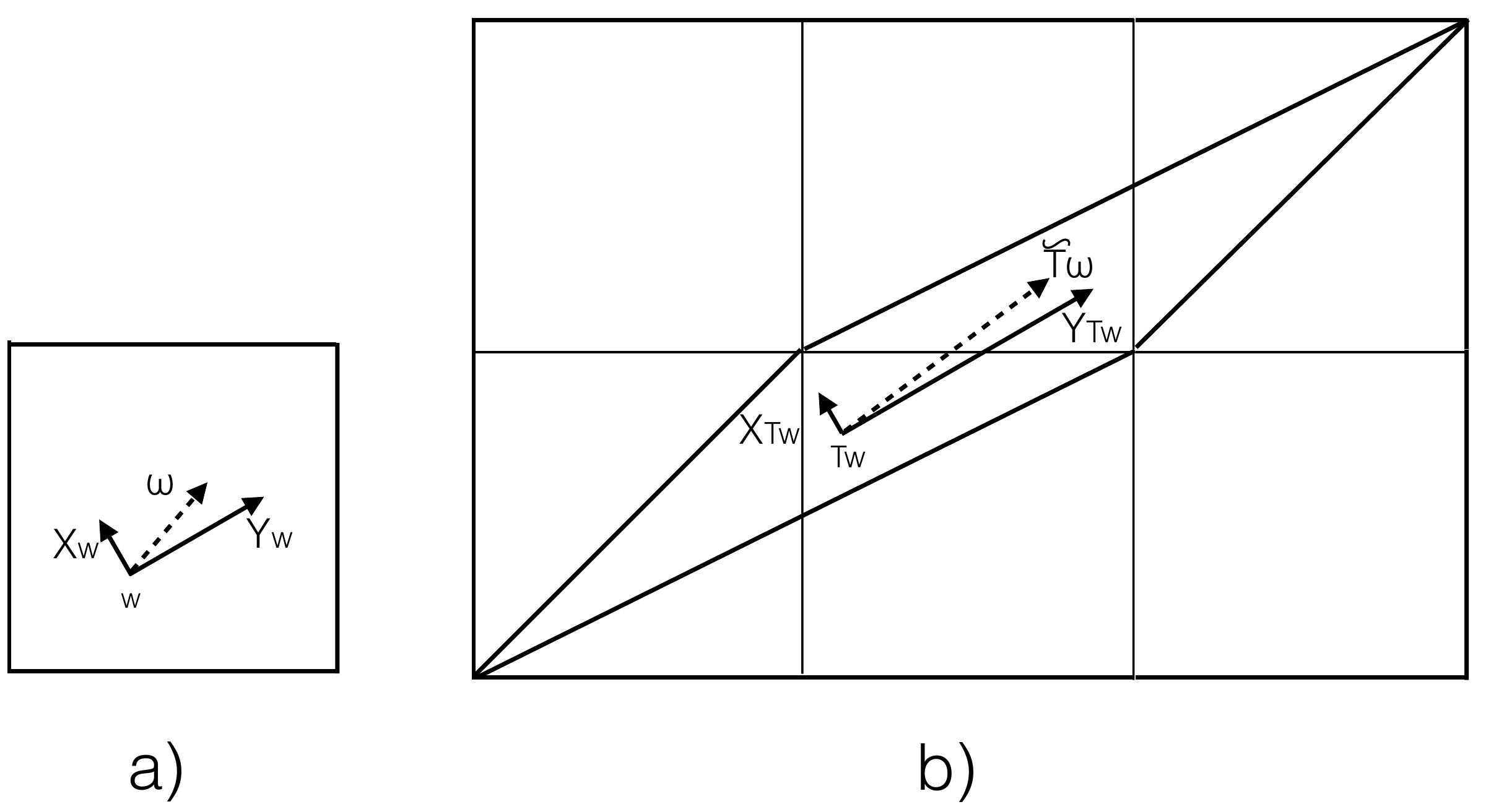}
\centering
\caption{The tangent vector $\omega \in R_{w}$ at 
the point $w \in W^2$  is decomposable  
into the sum $R_{w}= X_{w} \bigoplus Y_{w}$
where the spaces $X_w$ and $Y_w$ are defined by the corresponding 
eigenvectors of the $2 \times 2 $ matrix T (\ref{arno}).  
The automorphisms $T$ induces the mapping
of the tangent spaces $\tilde{T} X_{w} = X_{T w}, ~
\tilde{T} Y_{w} = Y_{T w}$.
It is contracting the distances on  $ X_{w}  $ and expanding the 
distances on  $ Y_{w} $.} 
\label{fig2}
\end{figure}
The C-condition requires that the tangent space $R^{m}_{w}$ at each point $w$
of the m-dimensional phase space $W^{m}$ of the dynamical system $\{T^{n}\}$ 
should be decomposable  into a 
direct sum of the two linear spaces  $X^{k}_{w}$ and $Y^{l}_{w}$ with the following 
properties \cite{anosov}:
\bea\label{ccondition}
C1.&R^{m}_{w}= X^{k}_{w} \bigoplus Y^{l}_{w} ~~\\
&The~ dynamical ~system~ \{T^{n}\}~ is ~such ~that: \nn\\
C2.&~~a) \vert \tilde{T}^{n} \xi  \vert  \leq ~ a \vert   \xi \vert e^{-c n}~ for ~n \geq 0 ; ~
\vert \tilde{T}^{n} \xi  \vert  \geq~ b \vert \xi \vert e^{-c n} ~for~ n \leq 0,~~~\xi \in  X^{k}_{w}, \nn\\
&b) \vert \tilde{T}^{n} \eta  \vert  \geq~ b \vert \eta \vert e^{c n} ~~for~ n \geq 0;~
\vert \tilde{T}^{n} \eta  \vert  \leq ~ a \vert   \eta \vert e^{c n}~ for~ n \leq 0,~~~\eta \in Y^{l}_{w},\nn
\eea
where the constants a,b and c are positive and are the same for all $w \in W^m$ and all 
$\xi \in  X^{k}_{w}$, $\eta \in Y^{l}_{w}$.
The length $\vert ...\vert$ of the tangent vectors   $\xi $ and $  \eta $  
is defined by the Riemannian metric $ds$ on $W^m$.

\begin{figure}
 \centering
\includegraphics[width=9cm]{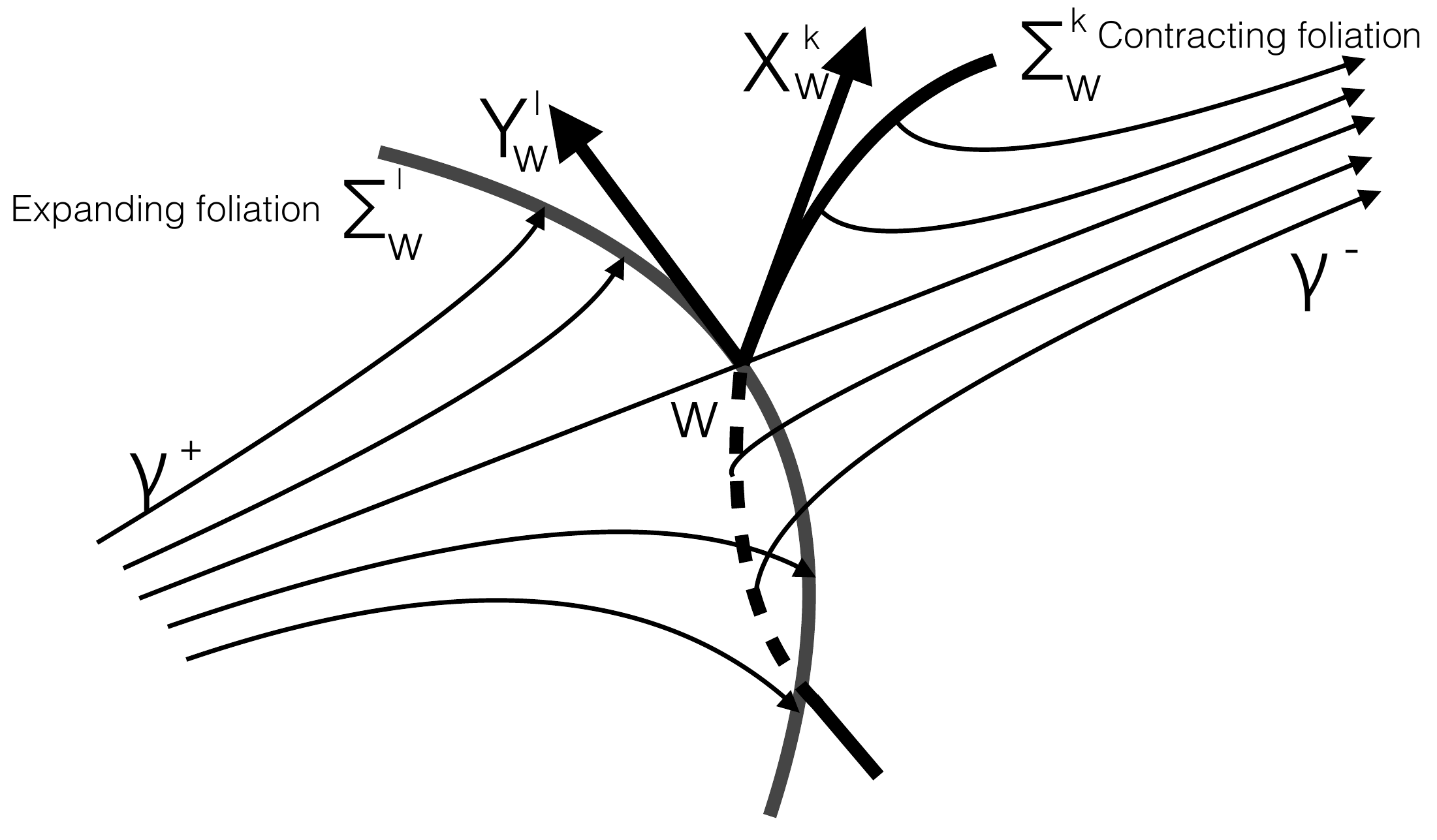}
\centering
\caption{ At each point $w$ of the C-system the tangent space $R^{m}_{w}$  
is  decomposable  into a direct sum of two linear spaces  $Y^l_{w} $ and $X^k_{w} $. 
The expanding and contracting geodesic flows
are $\gamma^+$ and $\gamma^-$. The expanding and 
contracting invariant  foliations  $\Sigma^l_{w} $ and $\Sigma^k_{w} $ 
{\it are transversal to the geodesic flows} and their corresponding tangent spaces 
are  $Y^l_{w} $ and $X^k_{w} $.  } 
\label{fig3}
\end{figure}

The linear spaces $X^{k}_{w}$ and $Y^{l}_{w}$ are invariant  with respect to 
the derivative  mapping  $\tilde{T}^{n} X^{k}_{w} = X^{k}_{T^n w}, ~
\tilde{T}^{n} Y^{l}_{w} = Y^{l}_{T^n w}$ and represent the {\it contracting and expanding 
linear spaces} (see Fig.\ref{fig2}).
The C-condition describes the behaviour of all trajectories $\tilde{T}^n \omega$ 
on the tangent vector bundle  $\omega \in R^{m}_{w}$. 
Anosov proved that the vector spaces  $X^{k}_{w}$ and $Y^{l}_{w}$ are continuous 
functions of the coordinate $w$ and that they are the target vector spaces  to 
the foliations $\Sigma^k$ and $\Sigma^l$ which are  the {\it  surfaces transversal to 
the trajectories}  $T^n w$ on $W^m$ (see Fig. \ref{fig3}). 
The contracting and expanding foliations  $\Sigma^k_w$ and $\Sigma^l_w$  are invariant 
with respect to the cascade $T^n$ in the sense that,  under the action of 
these transformations 
a foliation transforms into a foliation and, in general, into a different one \cite{anosov}. 

The interest in Anosov hyperbolic C-systems is associated with the earlier  
attempts to understand  relaxation phenomena of the hard scattering spheres and 
the foundation of the statistical mechanics by Krylov \cite{krilov}.  The hyperbolic systems 
demonstrate strong statistical properties and can help
with the understanding of the appearance of turbulence in fluid dynamics 
 \cite{turbul}, non-integrability of the Yang-Mills classical 
mechanics \cite{yangmillsmech,Savvidy:1982jk}, 
as well as dynamical properties of gravitating N-body 
systems and  the relaxation 
phenomena of stars in galaxies \cite{body,garry}. The present article is devoted to the
application of C-systems defined on a torus for the Monte-Carlo simulations of physical 
systems \cite{yer1986a}. The  random number generators based on C-systems 
demonstrated high quality statistical properties \cite{konstantin,hepforge} and are 
implemented  
into the ROOT and GEANT  software used at CERN for the LHC experiments \cite{cern,root,geant}.

In the next sections we shall consider the properties of the C-systems on a torus. 
The interest in the behaviour of the  C-systems on a torus is associated the with the fact that 
they can be used  to generate high quality pseudorandom numbers for 
the Monte-Carlo simulations of physical systems.  
In the second section we shall define a 
linear diffeomorphism on a torus which fulfils the C-condition and shall 
describe the distribution properties of their 
periodic trajectories. In the third section we shall define the Kolmogorov entropy of 
a dynamical system and calculate its value for the diffeomorphisms on a torus. 
In the fourth section we shall demonstrate that the integrals over the  phase space 
can be calculated as a sum over points of the infinitely long periodic trajectories. 
In the fifth section 
we shall present recent progress in the efficient implementation of a particular  C-system 
to generate pseudorandom numbers for Monte-Carlo simulations and 
their usage in high energy physics \cite{konstantin,hepforge,cern,root,geant}.  

\section{\it Torus Automorphisms}

Let us consider a dynamical system $T$ with the phase space $W^m$ 
which is the m-dimensional torus appearing at  factorisation of the 
Euclidean space $E^m$ with coordinates $ (w_1,...,w_m)$ over an integer lattice. 
Then $T$ can be thought of as a linear transformation of the Euclidean space $E^m$
which preserves the lattice $L$ of points with integer coordinates.
The automorphisms of the torus are generated by the linear transformation  
\bea\label{cmap}
w_i \rightarrow \sum T_{i,j} w_j,~~~~(mod ~1),
\eea
where the integer matrix $T$ has a determinant equal to one $Det T =1$. 
In order for the automorphisms of the torus (\ref{cmap})  {\it to fulfil  the C-condition it is necessary 
and sufficient that the matrix $T$ has no eigenvalues on the unit circle.}  
Thus the  spectrum $\{ \Lambda = {\lambda_1},...,
\lambda_m \}$ of the matrix $T$ is chosen to fulfil the following 
two conditions \cite{anosov}: 
\bea\label{mmatrix}
1)&~Det T=  {\lambda_1}{\lambda_2}....{\lambda_m}=1\nn\\
2)&\vert {\lambda_i} \vert \neq 1, ~~~~~~~~~~~\forall i.
\eea
Because the determinant of the matrix $T$ is equal to one,
the Liouville's measure $d\mu = dx_1...dx_m$ is invariant under the action of $T$.
The inverse matrix $T^{-1}$ is also an integer matrix because $Det T=1$.
Therefore $T$ is an automorphism of  the torus $W^m$  onto itself.
This automorphism has a fixed point p corresponding to the origin of $E^m$. 

\begin{figure}
 \centering
\includegraphics[width=7cm]{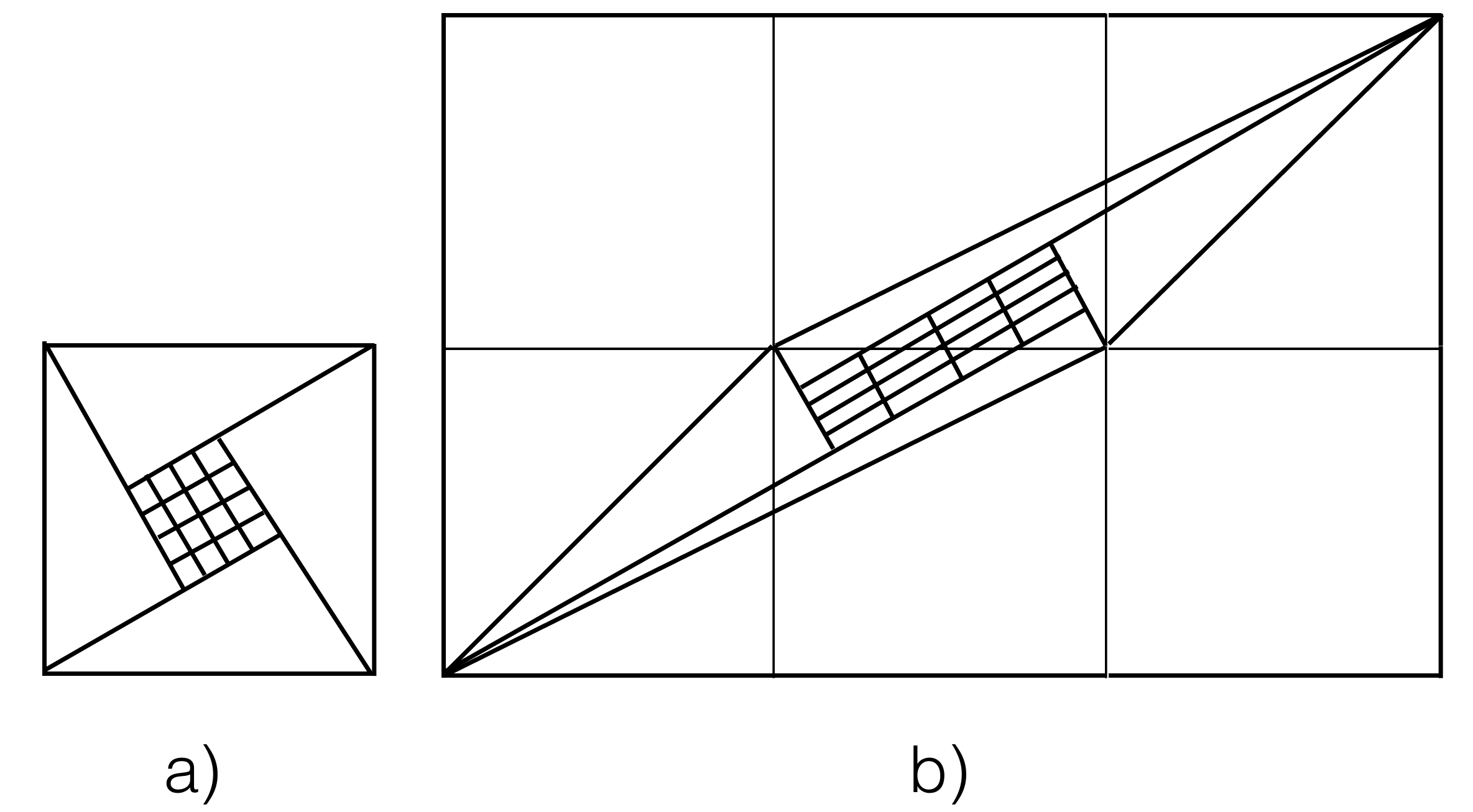}
\centering
\caption{The matrix $T$ (\ref{arno}) has two irrational  eigenvalues 
$ \lambda_{\alpha} = (3-\sqrt{5})/2$
and $\lambda_{\beta}= (3+\sqrt{5})/2$. The corresponding eigenvectors 
 $\{ e_{\alpha}  \}$ and  $\{ e_{\beta}  \}$ define 
two families of  parallel lines  $\{X_{\alpha} \}$ and $\{Y_{\beta} \}$ on the two-dimensional  
torus which are invariant under the automorphisms  $T$.
The automorphism $T$ is contracting  the distances between points on the 
lines belonging to the set $\{X_{\alpha} \}$ and expanding the 
distances between points on the lines belonging to the set $\{Y_{\beta} \}$.
The a) depicts the parallel lines of the sets  $\{X_{\alpha} \}$ and $\{Y_{\beta} \}$
and b) depicts their positions after the action of the automorphism  $T$.} 
\label{fig4}
\end{figure}

The above conditions (\ref{mmatrix}) on the eigenvalues of the matrix $T$ are  sufficient 
to prove that the system belongs to the class of  Anosov C-systems\footnote{If within the 
eigenvalues of the matrix T  there are {\it no roots of unity} 
then the automorphism T of a torus defines a K-system \cite{rokhlin1}. This means that 
some of the eigenvalues of T can be on a unit circle on the contrary to the 
 C-conditions (\ref{mmatrix}). 
It follows then that the K-systems occupy a different region in the space of dynamical 
systems $\{\dot{w}=f(w)\}$. In two and three dimensions the 
condition (\ref{mmatrix}) is equivalent to the absence of eigenvalues 
equal to the root of unity, thus in these cases the K- and C- conditions coincide, but in dimensions 
larger that three ($m \geq 4$) there are K-systems which are 
not C-systems \cite{anosov}. \label{note}} .
For that let us divide the eigenvalues of the matrix $T$ 
into the two sets   $\{ \lambda_{\alpha}  \} $ and $\{  \lambda_{\beta }  \} $ 
with modulus smaller and larger than one: 
\bea\label{eigenvalues}
0 <  \vert \lambda_{\alpha} \vert   < 1 <
\vert \lambda_{\beta}\vert. 
\eea
Consider  two  family of  planes $ X= \{X_{\alpha} \}$ and $ Y= \{Y_{\beta} \}$   
which are  parallel to the 
corresponding eigenvectors  $\{ e_{\alpha}  \}$ and  $\{ e_{\beta}  \}$ .
The derivative map  $\tilde{T}$ maps these planes into themselves  
contracting the points on the  $ X_{\alpha} $ plane  $\lambda_{\alpha}$ times  
and expanding  the points on the  $ Y_{\beta} $ plane  $\lambda_{\beta}$ times
(see Fig. \ref{fig4}):
\bea\label{expanding}
& \vert \tilde{T} \xi  \vert  \leq ~ \vert \lambda_{\alpha} \vert~   \vert   \xi \vert,~~~\xi \in  X_{\alpha}, \nn\\
&\vert \tilde{T} \eta  \vert  \geq~ \vert \lambda_{\beta}\vert~  \vert \eta \vert ,~~~\eta \in Y_{\beta}.
\eea
It follows then that  the automorphism T fulfils the C-condition (\ref{ccondition}) 
with a=b=1, $e^{c}= \lambda$, where 
\be\label{lowerbound}
\lambda = \min \left({1\over \max \vert \lambda_{\alpha} \vert}, \min\vert \lambda_{\beta}\vert \right).
\ee
In other words, 
the invariant planes of the matrix $T$,  for which the eigenvalues are inside or outside 
of the unit circle define the contracting and expanding invariant spaces 
of the C2-condition (\ref{ccondition}), so that
the phase trajectories of the dynamical system are expanding and contracting 
under the transformation $T$ at the exponential rate.  The foliations 
$\Sigma_{\alpha}$ and $\Sigma_{\beta}$
which are orthogonal  to the phase trajectories are divided into contracting  and 
expanding foliations and the tangent planes 
to the foliations  comprise the contracting and expanding   planes 
$ \{X_{\alpha} \}$ and $\{Y_{\beta} \}$. The foliations can be constructed
by filling out the Euclidean space $E^m$ by planes parallel to  
$\{X_{\alpha} \}$ and $\{Y_{\beta} \}$ and then projecting them  back to the torus $W^m$.   
 
Let us now consider the image of the set F of $W^m$ under the action 
of the transformation $T^n$.
It will be a hyperbolic rotation of F:  
expanding its points in the $e_{\beta}$ directions in $\vert \lambda_{\beta} \vert^n =
e^{n \ln \vert \lambda_{\beta}\vert}$ rate and 
contracting in the $e_{\alpha}$
directions in $\vert \lambda_{\alpha}\vert^n = e^{-n \ln{1\over \vert \lambda_{\alpha}\vert}}$ rate. 
Therefore at large n the image of the set $T^n F$ 
will be represented  by a long stripe stretched in the direction of the largest eigenvalue 
$\tilde{\lambda_{\beta}}$  and  uniformly distributed over the phase space $W^m$
in a way similar to the uniformly distributed 
trajectory of the equation $\dot{w} = \Omega$ on a torus with non-resonant 
vector $\Omega$\footnote{The vector $\Omega$ is non-resonant if the 
equation $n_1\Omega_1 +...+n_m \Omega_m =0$ has 
no solutions with integer $(n_1,...,n_m)$.}.  Thus, for large n,
the images $T^n F$ are crossing an arbitrary region $G \subset W^m$  
 proportional to its  volume $\mu(G)$:  
\be\label{mixing}
\lim_{n \rightarrow \infty} \mu(T^n F \cap G)= \mu(F) \mu(G).
\ee
This equation represents a {\it mixing} property of the dynamical system T. 
If one introduces a set of functions $\{ g(w) \}$ on the phase space $W^m$,
then one can represent the equation (\ref{mixing}) in the form 
\cite{kornfeld}\footnote{The equation (\ref{mixing}) follows from (\ref{funcform})
if one considers phase space functions which are equal to one on all points 
belonging to the  sets F and G 
and to zero for points outside the  sets.}:
\bea\label{funcform}
\lim_{n \rightarrow \infty} \int f(T^n w) ~g(w) d\mu(w)=  \int f(w) d\mu(w)
\int g(w) d\mu(w).
\eea
The above property is known in physics as the factorisation or decay of the 
two-point correlation functions. Because the C-systems have {\it mixing of 
all orders} \cite{anosov}  it follows that the C-systems are exhibiting  the decay of the 
 correlation functions of any order. 

A strong instability of trajectories of a dynamical C-system leads to the apperance 
of statistical properties in its behaviour \cite{leonov}.  As a result of exponential instability 
of the motion on the phase  space the time average  of the function $g(w)$
on $W^m$ 
\be\label{timeav}
{1 \over N} \sum^{N-1}_{n=0} g(T^n w)
\ee
behaves  as a superposition of quantities which are statistically weakly dependent. 
Therefore for the C-systems on a torus  it was  demonstrated that
the fluctuations of the time averages (\ref{timeav}) from the phase space integral 
\be\label{spaceav}
 \int_{W^m} g(w) d w 
 \ee
multiplied by $\sqrt{N}$ 
have at large $N \rightarrow \infty$ the Gaussian distribution \cite{leonov}:
\be\label{gauss}
\lim_{N\rightarrow \infty}\mu \bigg\{ w	:\sqrt{N}   \left({1 \over  N}  \sum^{N-1}_{n=0} g(T^n w)  - \int_{W^m} g(v)dv \right) < z    \bigg\}
= {1 \over \sqrt{2 \pi} \sigma_g}\int^{z}_{-\infty} e^{-{y^2 \over 2 \sigma^2_g}} dy.
\ee
The importance of the multiplication by the factor $\sqrt{N}$ can be understood as follows. 
The difference in the bracket has an upper bound in terms of  the Kolmogorov 
discrepancy $D_N(T)$ \cite{yer1986a,sobol}:
\be\label{descrepancy}
\bigg\vert   {1 \over  N}  \sum^{N-1}_{n=0} g(T^n w)  - \int_{W^m} g(v)dv       \bigg\vert 
 \leq C ~ {D_N(T) \over N},
\ee
where $C$ is a constant and $D_N(T)$ grows as $\sqrt{N}$. Therefore after multiplication of 
the quantity in the bracket by $\sqrt{N}$ it is bound by a constant.  The theorem 
states that the measure of points of $W^m$ which fulfil  the inequality (\ref{gauss})
have Gaussian  distribution.

Our next aim is to demonstrate  that C-systems have a rich variety  
of periodic trajectories, which essentially depends on the entropy of the C-systems
\cite{anosov,smale,sinai2,margulis,bowen0,bowen,bowen1}. 
We shall provide below a short introduction to the 
definition of the Kolmogorov entropy and shall calculate its value  for 
the different authomorphisms of a torus. 

\section{\it The Entropy of the Automorphisms T}

A  class of dynamical systems was introduced by Kolmogorov in \cite{kolmo,kolmo1}
which he called quasi-regular or K -systems and defined the notion of entropy 
for such systems.  Consider dynamical systems with discrete time.
Let $\alpha = \{A_i\}_{i \in I}$ ( $I$ is finite or countable)  be a  measurable partition of 
the phase space 
$W^m$ ($W^m$ is equipped with a positive measure $\mu$), that is 
\be
\mu(W^m \setminus \bigcup_{i \in I} A_i)=0,~~~~\mu(  A_i \bigcap   A_j)=0, i \neq j~,
\ee
then one can define the entropy of the partition $\alpha$ as 
\be
h(\alpha) = - \sum_{i \in I} \mu(A_i) \ln \mu(A_i).
\ee
It follows that if two partitions $\alpha_1$ and $\alpha_2$ differ by a set of 
zero measure then their entropies are equal.  The {\it refinement partition} $\alpha$ 
\be
\alpha = \alpha_1 \vee \alpha_2  \vee ... \vee \alpha_k
\ee
of the 
collection of partitions $\alpha_1,..., \alpha_k$ 
is defined as the intersection of all their composing sets  $A_i$:
\be
\alpha = \big\{ \bigcap_{i \in I} A_i~ \vert ~A_i \in \alpha_i~ for ~all~ i  \big\}.
\ee
The entropy of the partition $\alpha$ with respect to the automorphisms T 
is defined as a limit \cite{kolmo,kolmo1,sinai3,rokhlin1,rokhlin,rokhlin2}:
\be
h(\alpha, T)= \lim_{n \rightarrow \infty} {h(\alpha \vee T \alpha \vee ...\vee T^{n-1} \alpha) \over n},~~~~
n=1,2,...
\ee
This number is equal to the entropy of the refinement $\beta$: 
\be
\beta = \alpha \vee T \alpha \vee ...\vee T^{n-1} \alpha ,
\ee 
which was generated  during the iteration of the partition $\alpha$ 
by the  automorphism $T$. Finally the entropy of the 
automorphism $T$ is defined as a supremum: 
\be\label{supremum}
h(T) = \sup_{\{ \alpha \}} h(\alpha,T),
\ee
where the supremum is taken over all finite measurable partitions $\{ \alpha \}$ of  $W^m$.
From the definition it follows that  the calculation of this number for a 
given dynamical system seems extremely 
difficult. The theorem proven by Kolmogorov \cite{kolmo,kolmo1} tells that 
if one finds the so called "generating  partition" $\beta$ for the automorphisms $T$
which has the property\footnote{A sigma-algebra $\CM(\alpha)$ on $W^m$ is a family of subsets 
$\{ A_i \}_{i \in I} \in \alpha$ of the partition $\alpha$, which is closed under union  operations 
of countably many sets 
and   complement $\mu(\cup_{i\in I} A_i)= \sum_{i \in I}\mu(A_i)$.} 
\be
\bigcup^{\infty}_{n=-\infty} T^n ~\CM(\beta) = \CM,
\ee
where $\CM(\beta)$ is the sigma-algebra of the partition $\beta =  \{B_i\}_{i \in I}$ 
and $\CM$ is the sigma-algebra of all measurable sets of $W^m$, then
\be
h(T) = h(\beta,T).
\ee
Which means that the supremum in  (\ref{supremum}) is actually reached on a generating partition 
$\beta$. In some cases the construction of the generating partition $\beta$
allows an explicit  calculation of the entropy of a given dynamical system \cite{sinai4,gines}. 

There are also alternative ways of  calculating the entropy of a C-system. The most convenient 
for us is the integration over the whole 
phase space of the logarithm of the volume expansion rate $\lambda(w)$ 
of a $l$-dimensional infinitesimal cube which is embedded  into the foliation  $\Sigma^{l}_w$. 
The derivative map
$\tilde{T}$ maps the linear space $Y^{l}_{w}$ into the $Y^{l}_{Tw}$ and if 
the rate of expansion of the volume of the $l$-dimensional cube is  $\lambda(w)$,
then \cite{anosov,sinai3,rokhlin2,sinai4,gines}
\be\label{biuty}
h(T) = \int_{W^m} \ln \lambda(w) d w.
\ee
Here the volume of the $W^m$ is normalised to 1. For the automorphisms 
on a torus (\ref{cmap}) the coefficient $\lambda(w)$  does not depends of the phase 
space coordinates $w$ and is equal to the product of eigenvalues 
$\{  \lambda_{\beta }  \} $  with modulus  larger than one (\ref{eigenvalues}): 
\be\label{more}
\lambda(w) = \prod^{l}_{\beta=1} \lambda_{\beta}.
\ee
Thus for the  Anosov automorphisms on a torus (\ref{cmap}), (\ref{mmatrix})
from (\ref{biuty}) and (\ref{more}) we can calculate the entropy, which became  
equal to the sum:
\be\label{entropyofT}
h(T) = \sum_{\vert \lambda_{\beta} \vert > 1} \ln \vert \lambda_{\beta} \vert.
\ee

It is interesting to know  what will happen if 
the authomorphism of a torus is defined by the k'th power $T^k$ of the matrix $T$. 
In that case it will generate a cascade of the form $\{ (T^k)^n\}$ and the question is what 
is the entropy of this authomorphism. The answer can be found by calculating  the 
eigenvalues of the new matrix, therefore
\be
h(T^k) = \sum_{\vert \lambda_{\beta} \vert > 1} \ln \vert \lambda_{\beta} \vert^k = 
k \sum_{\vert \lambda_{\beta} \vert > 1} \ln \vert \lambda_{\beta} \vert = k ~h(T).
\ee 
As one can see, the entropy increases linearly with the power of the matrix $k$ and, as we shall 
see in the next sections, this is a very "expensive" and time consuming way of increasing 
the entropy of the C-systems used for Monte-Carlo calculations.

In the next section we shall present the alternative derivation of the above results,
which will demonstrate 
the connection of the entropy with the variety  and richness of the periodic 
trajectories of the C-systems  \cite{anosov,bowen0,bowen,bowen1}.

\section{\it Periodic Trajectories  of Anosov C-systems}

Let us demonstrate now that the C-systems have a countable set of 
periodic trajectories \cite{anosov}.
The $E^m$ cover of the torus $W^m$ allows 
to translate every set of points on torus into a set of points on Euclidean space $E^m$ and 
the space of functions on torus into the periodic functions on $E^m$.
To every closed curve $\gamma$ on a torus corresponds a curve  $\phi: [0,1] \rightarrow E^m$
for which $\phi(0) = \phi(1)~ mod~1$ and if $\phi(1) - \phi(0)=(p_1,...,p_m)$, then the 
corresponding winding numbers on a torus are $p_i \in Z$. 

{\it All trajectories with rational coordinates $(w_1,...,w_m)$, and only they, 
are periodic trajectories of the automorphisms of the  torus} (\ref{cmap}).
Let us fix the integer number $N$, then the points on a torus with 
the coordinates having a denominator $N$ form a finite set $\{p_1/N,...,p_m/N \}$. The 
automorphism (\ref{mmatrix}) with integer entries transform this set 
of points into itself, therefore all these points belong to periodic trajectories.
Let  $w=(w_1,...,w_m)$ be a point of a trajectory with the period $n > 1$. Then 
 \be\label{periodictrajectories1}
 T^n w = w + p,
 \ee
 where $p$ is an integer vector. The above equation with 
 respect to $w$ has nonzero determinant, therefore the components 
 of $w$ are {\it rational}. 
 
Thus the periodic trajectories of the period $n$ of the automorphism $T$ are given 
by the solution of the equation (\ref{periodictrajectories1}),
where $p \in Z^m$ is an integer vector and $w=(w_1,...,w_m) \in W^m$.  
 As $p$ varies in $Z^m$ the solutions of the equation (\ref{periodictrajectories1}) determine
 a fundamental domain $D_n$ in the covering Euclidian space $E^m$  of the volume 
 $\mu(D_n)=1/\vert Det (T^n-1) \vert $. Therefore the number of all points $N_n$ on the 
 periodic trajectories of the period $n$  
 is given by  the corresponding inverse volume \cite{smale,sinai2,margulis,bowen0,bowen}: 
 \be\label{numbers}
N_n = \vert Det (T^n-1) \vert = \vert  \prod^{m}_{i=1}(\lambda^n_i -1) \vert .
 \ee
Using the theorem of Bowen \cite{bowen,bowen1} which states that the entropy  
of the automorphism $T$ can be  represented  in terms of $N_n$
defined in   (\ref{numbers}): 
\be\label{bowen}
h(T)= \lim_{n \rightarrow \infty}{1\over n}~ \ln  N_n ~~,  
\ee
one can derive the formula for the entropy (\ref{entropyofT}) for the automorphism $T$
in terms of its eigenvalues: 
\be
h(T)= \lim_{n \rightarrow \infty}{1\over n} \ln (\vert  \prod^{m}_{i=1}(\lambda^n_i -1) \vert) =
\sum_{\vert \lambda_{\beta} \vert > 1} \ln \vert \lambda_{\beta} \vert.
\ee
Let us now define the number of periodic trajectories of the period  $n$  by $\pi(n)$.
Then the number of all points $N_n$ on the 
 periodic trajectories of the period $n$  can be written in the following form:
 \be
 N_n = \sum_{l~ divi~ n  } l ~\pi(l) ,
 \ee
where $l$ divides $n$. Using again the Bowen result (\ref{bowen}) one can get 
\be\label{assimpto}
N_n =\sum_{l~ divi~ n  } l \pi(l) \sim e^{n h(T)}.
\ee
This result can be rephrased as a statement that the number of points 
on the periodic trajectories of the period n exponentially 
grows  with the entropy.

Excluding the periodic trajectories which divide n 
(for example $T^n w=T^{l_2}(T^{l_1}w)$, where $n=l_1 l_2$ and $T^{l_i}w =w$)  
one can get the number of periodic trajectories of period n 
which are not divisible.
For that one should represent  the $\pi(n)$ in the following form:
\be\label{density}
\pi(n) = {1\over n} \big( \sum_{l~ divi~ n  } l ~\pi(l) -
\sum_{l~ divi~ n,~ l <n  } l ~\pi(l) \big)
\ee
and from (\ref{density})  and (\ref{assimpto}) it follows that 
\be
\pi(n) \sim { e^{n h(T)} \over n}  \big( 1   -
{\sum_{l~ divi~ n,~ l <n  } l ~\pi(l) \over \sum_{l~ divi~ n  } l ~\pi(l)} \big) \sim { e^{n h(T)} \over n}, 
\ee
because  the ratio in the bracket is strictly smaller than one. This result tells that a
system with larger entropy  $\Delta h = h(T_1) - h(T_2) >0$ is more densely populated by the 
periodic trajectories of the same period $n$:
\be
{\pi_1(n) \over \pi_2(n)} \sim e^{n\ \Delta h}.
\ee
The next important result of the Bowen theorem \cite{bowen,bowen1} states that 
\be\label{ation9}
\int_{W^m} f(w) d\mu(w) = \lim_{n \rightarrow \infty}  {1\over N_n} \sum_{ w \in \Gamma_n} f(w),
\ee
where $\Gamma_n$ is a set of all points on the trajectories of period
$n$. 
The total number of points in the set $\Gamma_n$ we defined earlier as $N_n$.

This result has important consequences for the calculation of the 
integrals on the manifold $W^m$, because, as it follows from (\ref{ation9}), the integration 
reduces to the summation 
over all points of periodic trajectories. It is appealing to consider periodic trajectories 
of the period $n$ which is a prime number. Because every infinite subsequence of convergent  
sequence converges to the same limit we can consider in (\ref{ation9}) only terms 
with the prime periods.  In that case $N_n = n \pi(n)$ and the above formula becomes:
\be\label{integral}
\int_{W^m} f(w) d\mu(w) = \lim_{n \rightarrow \infty}  {1\over n \pi(n)} \sum^{\pi(n)}_{j=1}
\sum^{n-1}_{ i=0} f(T^i w_{j}),
\ee
where the summation is over all points of the trajectory $T^i w_{j}$ and  over all 
distinct  trajectories of period n which are enumerated by index $j$. The $w_{j}$ is the initial point of the 
trajectory $j$ \footnote{It appears to be a difficult mathematical problem to decide whether two  vectors 
$w_{1}$ and $w_{2}$ belong to the same or to distinct trajectories.}.  
From the above consideration it follows that the convergence is guaranteed 
if one sums over all trajectories of the same period $n$.    One can conjecture     
that all $\pi(n)$ trajectories at the very large period $n$  contribute  
equally into the sum (\ref{integral}),  
therefore the integral (\ref{integral}) can be reduced to a  sum over fixed trajectory
\be\label{reduce}
  {1\over n } \sum^{n-1}_{ i=0} f(T^i w).
\ee
This reduction seems plausible, but even in that case the calculation of  (\ref{reduce}) 
is impractical, because the periods  of the  C-systems which are used 
for Monte-Carlo calculations can be as large as $n \sim 10^{4000}$ \cite{konstantin}.
The Monte-Carlo calculation of the integrals is  therefore  performed on some parts
of the length  $N$ of the long trajectories  \cite{metropolis,neuman,neuman1,sobol}. In that case 
the approximation of the integrals is regulated by the Kolmogorov discrepancy 
$D_N(T)$ in the formula (\ref{descrepancy}).  The behaviour of $D_N(T)$ 
as a function of $N$ crucially 
depends on the properties of the dynamical system $T$ because  the discrepancy grows 
slower for the C-systems  periodic trajectories which are distributed uniformly 
and everywhere dense in the phase space $W^{m}$ \cite{anosov,yer1986a}. 
Therefore  it was suggested that  
Anosov C-systems \cite{anosov},  defined on a high dimensional 
torus, are excellent candidates for the pseudo-random number generators \cite{yer1986a}.
In the next section we shall demonstrate how these results can be used for 
practical calculations and especially in high energy particle physics.

\section{\it  Anosov C-systems for Monte-Carlo Computations }

Modern powerful computers open a new era for the application of the Monte-Carlo Method 
 \cite{metropolis,neuman,neuman1,sobol,yer1986a,fred,Demchik:2010fd,falcion} 
 for the simulation of physical systems with many degrees of freedom and of 
higher complexity. The Monte-Carlo simulation is an important computational 
technique in many areas of natural sciences, and it has significant application 
in particle and nuclear physics, quantum physics, statistical physics, 
quantum chemistry, material science, among many other multidisciplinary applications. 
At the heart of the Monte-Carlo (MC) simulations are pseudo Random Number Generators (RNG).

Usually  pseudo random numbers are generated by deterministic recursive rules
\cite{yer1986a,metropolis,neuman,neuman1,sobol}. 
Such rules produce pseudorandom numbers, and it is a great challenge to design 
pseudo random number generators that produce high quality sequences. 
Although numerous RNGs introduced in the last  decades fulfil most of the 
requirements and are frequently used in simulations, each of them has some 
weak properties which influence the results \cite{pierr} and are less suitable for demanding 
MC simulations which are performed for the high energy experiments at CERN
and other research centres. 
The RNGs are essentially used in ROOT and GEANT software for the design of the experiments  
and statistical analysis of the experimental data \cite{cern}. 

In order to fulfil these demanding requirements it is necessary to have a solid theoretical 
and mathematical background on which the RNG's are based. RNG  should have a long period, 
be statistically robust, efficient, portable and have a possibility to change and 
adjust the internal characteristics in order to make RNG suitable for concrete 
problems of high complexity.

In \cite{yer1986a} it was suggested that  
Anosov C-systems \cite{anosov},  defined on a high dimensional 
torus, are excellent candidates for the pseudo-random number generators.
The C-system chosen in \cite{yer1986a} was the one which realises 
linear automorphism  $T$  defined  in (\ref{cmap})  \footnote{For convenience 
in this section the dimension $m$ of the phase space $W^m$ is denoted by  $N$.}:
\bea\label{cmap1}
w_i \rightarrow \sum^{N}_{j=1} T_{i,j} w_j,~~~~(mod ~1).
\eea
A particular matrix chosen in \cite{yer1986b} was defined for all $N \geq 3$:
\be
\label{eqmatrix}
T =  
   \begin{pmatrix}  
      1 & 1 & 1 & 1 & ... &1& 1 \\
      1 & 2 & 1 & 1 & ... &1& 1 \\
      1 & 3+s & 2 & 1 & ... &1& 1 \\
      1 & 4 & 3 & 2 &   ... &1& 1 \\
      &&&...&&&\\
      1 & N & N-1 &  N-2 & ... & 3 & 2
   \end{pmatrix}
\ee
The matrix $T$ of integer numbers is of the size $N \times N$ and has 
determinant equal to one. It is defined recursively, 
since the matrix of size $N+1$ contains in it the matrix of the size $N$.  
The only variable entry in the matrix is $T_{32} =3+s$, where  $s$  
should be chosen such that to avoid eigenvalues lying on a unit circle.
In order to generate pseudo-random vectors  $w_n = T^n w$, one should 
 choose  the initial vector $w=(w_{1},...,w_{m})$, called the ``seed",  
 with at least one non-zero component to avoid fixed point of $T$, which is at the origin.

\begin{figure}[htbp]
   \centering
   \includegraphics[width=14cm]{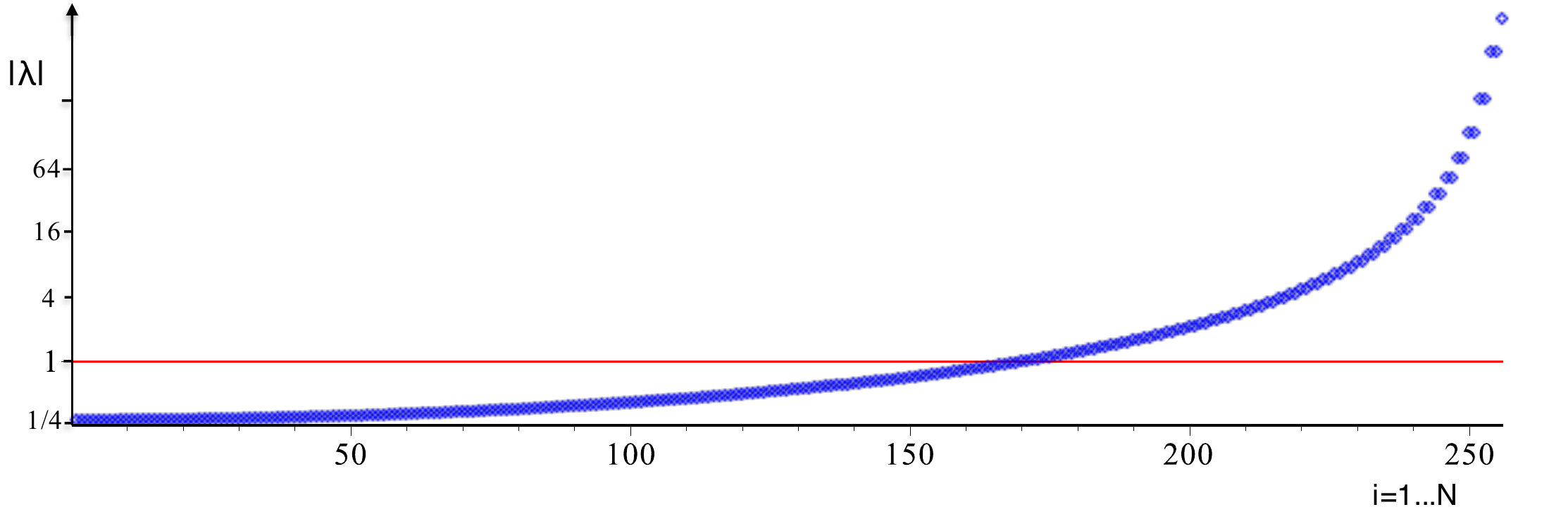}  
   \caption{The logariphm of the absolute value of the eigenvalues of the MIXMAX 
   matrix for the size N=256. The area under the curve is proportional 
   to the entropy (\ref{entropyofT}).}
   \label{fig:mixmaxev}
\end{figure}

The eigenvalues of the $T$ matrix (\ref{eqmatrix}) are widely dispersed for all $N$,
see Fig.  \ref{fig:mixmaxev}
from reference \cite{konstantin}. The spectrum  is "multi-scale", with trajectories exhibiting exponential instabilities at different scales \cite{yer1986a}. The entropy of $T$  
satisfies the  bound \cite{konstantin}:
\be
{4\over 9}\,N \, \ln 4  <    h   <    N \ln 4 ,
\ee
and tells that the entropy of $T$ is {\it increasing linearly with its size $N$}. In the light of the discussion 
in the previous section this means that there is a rich set of periodic trajectories 
with large periods and their  variety  {\it increases  exponentially with the size $N$} of the matrix $T$.  
For  the matrix of the size $N=256$ the actual value is $h \simeq 164.4$ \cite{konstantin}, thus 
\be
\pi(n) \sim {e^{164\ n}\over n}.
\ee
The Monte-Carlo simulations of a Markov chain should be done with a random number generator whose auto-correlation time $\tau_0 = 1/h$ is much smaller than the auto-correlation time of the Markov chain \cite{yer1986a}.

In a typical computer implementation of the authomorphism \eqref{cmap1} 
the initial vector will have rational 
components $u_i=a_i/p$, where $a_i$ and $p$ are natural numbers.  
Therefore it is convenient to represent $u_i$ by its numerator $a_i$ in computer memory and define the iteration in terms of $a_i$
\cite{mixmaxGalois}:
\be
\label{eq:recP}
a_i \rightarrow\sum_{j=1}^N T_{ij} \, a_j ~\textrm{mod}~ p .
\ee
If the denominator p is taken to be a prime number \cite{mixmaxGalois}, 
then the recursion is realised on extended 
Galois field $GF[p^N]$  \cite{niki,nied} and 
allows to find the period of the trajectory n in terms of p and the properties of the 
characteristic polynomial $P(x)$ of the matrix T \cite{mixmaxGalois}. If 
the characteristic polynomial $P(x)$ of some matrix $T$ is primitive in the 
extended Galois  field $GF[p^N]$, then
\cite{mixmaxGalois,nied,lnbook}
\be\label{period}
 T^q = p_0~ \mathbb{I}~~\textrm{ where}~~  q=\frac{p^N-1} {p-1} ~,
\ee
where $p_0$ is a free term of the  polynomial $ P(x)$ and is a {\it primitive element} of $GF[p]$.
Since our matrix T has $p_0=Det T= 1$, the polynomial $ P(x)$ of T cannot be primitive. 
The solution suggested  in \cite{konstantin} is to define the necessary and 
sufficient conditions for the period $q$  
to attain its maximum, they are:
\begin{enumerate}
\item[\bf{1.}] $T^q = \mathbb{I} ~(mod~ p) $,~~~where $q=\frac{p^N-1} {p-1}$
\item[\bf{2.}] $T^{q/r} \neq \mathbb{I} ~(mod~ p)$,~~~~ for any r which is a prime divisor of q .
\end{enumerate}
The first condition is equivalent to the requirement  that the characteristic polynomial is irreducible. 
The second condition can be checked if the integer factorisation of $q$ is available \cite{konstantin}, then
the period of the sequence is equal to (\ref{period}) and is independent of the seed. 
There are precisely $p-1$ distinct  trajectories which together fill up all states of the $GF[p^N]$ lattice: 
\be
 q~ (p-1) = p^N-1.
\ee
In \cite{konstantin} the actual value of p was  taken as $p=2^{61}-1$, 
the largest Mersenne number that fits into an 
unsigned integer on current 64-bit computer architectures.  For the matrix of the 
size $N=256$ the period in that case is  $q \approx 10^{4600}$.

The algorithm  which allows the efficient implementation of the generator 
in actual computer hardware, reducing the matrix multiplication to the O(N) operations
was found in \cite{konstantin}.
The other advantage of this implementation is that it allows to make  "jumps" into 
any point on a periodic trajectory 
without calculating all previous coordinates on a trajectory, which typically has a 
very large  period $q \approx 10^{4600}$.
This MIXMAX random number generator is currently made available 
in  a portable implementation  in the C language at hepforge.org \cite{hepforge}
and was implemented into the ROOT and GEANT projects at CERN \cite{cern,root,geant}.

\section{\it Conclusion}

The author had the privilege to visit Professor Dimitry Anosov at the 
Steklov Institute in 1986 to discuss and  learn  his fundamental work on 
hyperbolic dynamical systems.  Professor Anosov's impressive  simplicity and clear 
explanation of complicated mathematical constructions was an unforgettable experience. 

\section{\it Acknowledgement }
I would like to thank L. Moneta, J.Apostolakis, J.Harvay, H.Asatryan, H.Babujyan, 
R.Poghosyan, R.Manvelyan and K.Savvidy  for stimulating discussions, 
as well as the  CERN PH-SFT department 
and A. Alikhanian National Laboratory in Yerevan, where part of this work was completed, 
for kind hospitality. This work was supported in part by the European Union's Horizon 2020 
research and innovation programme under the Marie Sk\'lodowska-Curie 
grant agreement No 644121.

\section{\it Note Added}

\begin{figure}
 \centering
\includegraphics[width=7cm]{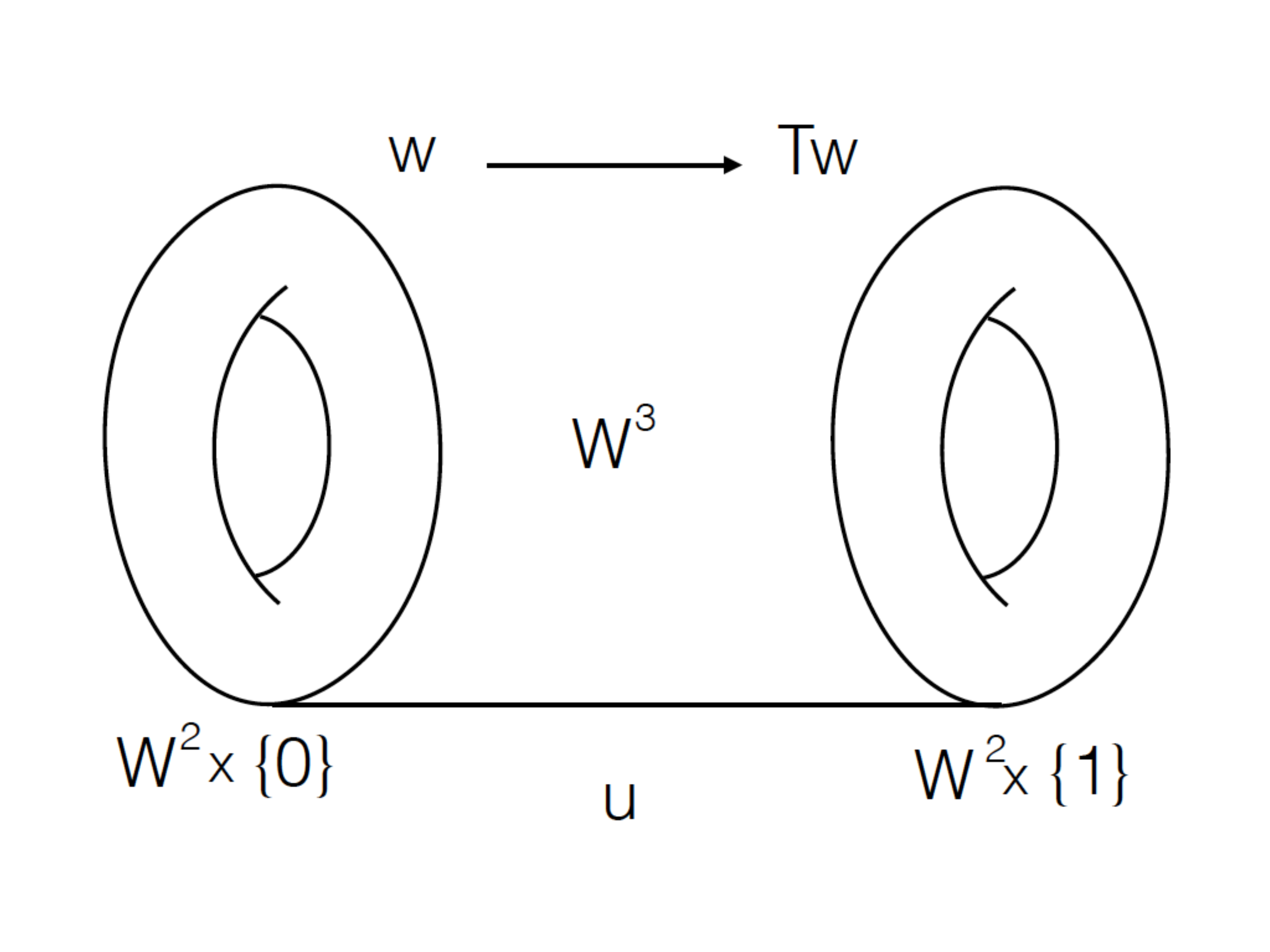}
\caption{The  identification of the  $W^2 \times \{0\}$ with $W^2 \times \{1\}$ by the 
formula $(w,1) \equiv  (Tw,0)$ of a cylinder $W^2 \times [0,1]$, where $[0,1]= 
\{  u~ \vert ~ 0 \leq u \leq 1 \}$.
The resulting compact manifold $W^{3}$ has a bundle structure 
with the base $S^1$ and fibres of the type $W^2$. The manifold $W^{3}$ 
has the local coordinates $\tilde{w}=(w^1,w^2,u)$ . 
} 
\label{fig6}
\end{figure}
In \cite{anosov} Anosov demonstrated how any C-cascade on a torus can be
embedded into a certain  C-flow. The embedding was defined by the identification 
(\ref{identification}) and the corresponding  C-flow on a smooth Riemannian manifold 
$W^{m+1}$ with the metric (\ref{metric}) was defined by the equations (\ref{velo}).
We are interested here to analyse the {\it geodesic flow} on the same Riemannian manifold $W^{m+1}$. 
The geodesic flow has different dynamics (\ref{geodesicflow} ) and  as we shall demonstrate below 
has very interesting hyperbolic components different from (\ref{velo}). 

Let us consider a C-cascade on a torus $W^m$
(defined in section two) and increase its dimension m  by one unit
constructing a cylinder $W^m \times [0,1]$, where $[0,1]= \{  u~ \vert ~0 \leq u \leq 1 \}$,
and identifying $W^m \times \{0\}$ with $W^m \times \{1\}$ by the formula:
\be\label{identification}
(w,1) \equiv  (Tw,0).
\ee
Here T is diffeomorphism (\ref{cmap}):
\bea\label{cmap1}
 w^i \rightarrow \sum T_{i,j} w^j,~~~~(mod ~1).
\eea
The resulting compact Riemannian manifold $W^{m+1}$ has a bundle structure
with the base $S^1$ and fibres of the type $W^m$. The manifold $W^{m+1}$
has the local coordinates $\tilde{w}=(w^1,...,w^m,u)$ .

The C-flow $T^t$ on the manifold $W^{m+1}$ is defined by the equations \cite{anosov}
\be\label{velo}
{d  w^1 \over d t}=0~, ....,~ {d  w^m  \over d t} = 0,~ {d  u \over d t}=1.
\ee
For this flow the tangent space $R^{m+1}_{\tilde{w}}$ can be represented as
a direct sum of three subspaces - contracting  and  expanding
 linear spaces  $X^k_{\tilde{w}} $,$ Y^l_{\tilde{w}} $   and $ Z_{\tilde{w}}$:
 \be
R^{m+1}_{\tilde{w}} = X^k_{\tilde{w}}  \oplus  Y^l_{\tilde{w}}  \oplus  Z_{\tilde{w}}.
\ee
The linear space  $X^k_{\tilde{w}} $ is tangent to the fibre  $W^m \times u$ and is parallel
to the eigenvectors corresponding to the eigenvalues which are lying inside the unit circle  $0 <  \vert \lambda_{\alpha} \vert   < 1$ and $Y^l_{\tilde{w}} $ is tangent to the fibre  $W^m \times u$ and is parallel to the eigenvectors  corresponding to the eigenvalues
which are lying outside of the unit circle $1 <\vert \lambda_{\beta}\vert$. $ Z_{\tilde{w}}$ is collinear to the phase space velocity (\ref{velo}). Under the derivative mapping of the (\ref{velo}) the vectors (\ref{eigenvectros}) from $X^k_{\tilde{w}} $ and  $Y^l_{\tilde{w}} $ are contracting  and
stretching:
\be
\vert \tilde{T}^{t} v_1 \vert =  \lambda_2^{ t}~ \vert v_1 \vert,~~~~
\vert \tilde{T}^{t} v_2 \vert =  \lambda_1^{ t}~ \vert v_2 \vert.
\ee
This identification of contracting and expanding spaces proves  that (\ref{velo})
indeed defines a C-flow \cite{anosov}.

It is also interesting to analyse the {\it geodesic flow} on a  Riemannian manifold
$V^n \equiv W^{m+1}$ (n=m+1). The
equations for the geodesic flow on $V^{n}$
\be
{d^2 \tilde{w}^{\mu} \over d t^2} +\Gamma^{\mu}_{\nu \rho}
{d  \tilde{w}^{\nu} \over d t } {d  \tilde{w}^{\rho} \over d t } =0
\ee
are different from the flow equations
defined by the equations (\ref{velo}) and our goal is to learn if the geodesic flow
has also the properties of the C-flow. The answer to this  question is not obvious and
requires investigation of the curvature structure of the manifold $V^{n}$. If all sectional
curvatures on $V^{n}$ are negative then geodesic flow defines a C-flow \cite{anosov}.

For simplicity we shall consider a  two-dimensional case which is
defined by the matrix
\be\label{arno}
T= \begin{pmatrix}
1&1\\
1&2\\
 \end{pmatrix} .
 \ee
The metric on the corresponding manifold $V^3$ can be defined as \cite{arnoldavez}
\be\label{metric}
ds^2 = e^{2u} [\lambda_1 d w^1 + (1-\lambda_1) d w^2]^2 +
e^{2u} [\lambda_2 d w^1 + (1-\lambda_2) d w^2]^2 +du^2 =\nn\\
 g_{\mu\nu} d\tilde{w}^{\mu} d\tilde{w}^{\nu},
\ee
where $0 < \lambda_2  < 1 <  \lambda_1$ are eigenvalues of the matrix (\ref{arno}) and
fulfil the relations $\lambda_1 \lambda_2 =1,\lambda_1+ \lambda_2=3$.
The metric  is invariant  under the transformation
\be\label{trans}
w^1 = 2 w^{'1} - w^{'2},~~~ w^2 = -w^{'1}_1 + w^{'2}, ~~~u  = u^{'}-1
\ee
and is therefore consistent with the identification (\ref{identification}). The metric
tensor has the form
\be\label{metric}
 g_{\mu\nu}(u)= \begin{pmatrix}
\lambda_1^{2 + 2 u} + \lambda_2^{2 + 2 u} & (1 - \lambda_1) \lambda_1^{1 + 2 u} + (1 - \lambda_2) \lambda_2^{1 + 2 u}& 0 \\
(1 - \lambda_1)\lambda_1^{1 + 2 u} + (1 - \lambda_2) \lambda_2^{1 + 2 u}&(1 - \lambda_1)^2 \lambda_1^{2 u} + (1 - \lambda_2)^2 \lambda_2^{2 u}&0\\
0&0&1\\
 \end{pmatrix}
 \ee
and the corresponding geodesic equations take the following form:
 \bea\label{geodesicflow}
&  \ddot{w}^1 + 2{(\lambda_1-1) \ln\lambda_1 \over \lambda_1+1} \dot{w^1}\dot{u}
-4 {(\lambda_1-1) \ln\lambda_1 \over \lambda_1+1} \dot{w^2}\dot{u}  =0
\nn\\
&  \ddot{w}^2 - 2{(\lambda_1-1) \ln\lambda_1 \over \lambda_1+1} \dot{w^2}\dot{u}
  - 4 {(\lambda_1-1) \ln\lambda_1 \over \lambda_1+1} \dot{w^1}\dot{u} =0\\
& \ddot{u} + {(1-\lambda^{4u+4}_1) \ln\lambda_1 \over \lambda^{2u+2}_1} \dot{w^1}\dot{w^1}
 + 2{(1+\lambda^{4u+3}_1)(\lambda_1-1) \ln\lambda_1 \over \lambda^{2u+2}_1} \dot{w^1}\dot{w^2}+\nn\\
& +  {(1-\lambda^{4u+2}_1)(\lambda_1-1)^2 \ln\lambda_1 \over \lambda^{2u+2}_1} \dot{w^2}\dot{w^2}=0.\nn
\eea
One can become convinced that these equations are invariant under the transformation (\ref{trans}).
 In order to study a stability of the geodesic flow one has to compute the sectional curvatures.
We shall choose the orthogonal frame in the directions of the linear spaces  $X^1_{\tilde{w}} , Y^1_{\tilde{w}} $   and $ Z_{\tilde{w}}$. The corresponding vectors are:
\be\label{eigenvectros}
v_1 = (\lambda_1-1, \lambda_1,0),~~~v_2=(\lambda_2 -1, \lambda_2,0),~~~ v_3=(0,0,1)
\ee
and in the metric (\ref{metric}) they have the lengths:
\be
\vert v_1  \vert^2= (\lambda_1 - \lambda_2)^2 \lambda_2^{2 u},~~~~
\vert v_2  \vert^2= (\lambda_1 - \lambda_2)^2 \lambda_1^{2 u},~~~~
\vert v_3 \vert^2 = 1.
\ee
The corresponding sectional curvatures are:
\bea
K_{12} = {R_{\mu\nu\lambda\rho} v^{\mu}_1 v^{\nu}_2  v^{\lambda}_1 v^{\rho}_2 \over
\vert v_1 \wedge v_2 \vert^2}=   \ln^2 \lambda_1
\nn\\
K_{13} = {R_{\mu\nu\lambda\rho} v^{\mu}_1 v^{\nu}_3  v^{\lambda}_1 v^{\rho}_3 \over
\vert v_1 \wedge v_3\vert^2}= -  \ln^2 \lambda_2
\\
K_{23} = {R_{\mu\nu\lambda\rho} v^{\mu}_2 v^{\nu}_3  v^{\lambda}_2 v^{\rho}_3 \over
\vert v_2 \wedge v_3 \vert^2}= -  \ln^2 \lambda_1. \nn
\eea
It follows for the above equations that the geodesic
flow is exponentially unstable on the planes (1,3) and (2,3)
and is stable in the plane (1,2). This behaviour is dual to the flow (\ref{velo}) which
is unstable in (1,2) plane and is stable in (1,3) and (2,3) planes.
 The scalar curvature is
\be
R= R_{\mu\nu\lambda\rho} g^{\mu\lambda}g ^{\nu\rho} = 2(K_{12} +K_{13}+K_{23}) =   - 2 
\ln^2 \lambda_1 = -2 h(T)^2 ,
\ee
where $h(T)$ is the entropy of the automorphism T (\ref{arno}).

\vfill


\begin{thebibliography}{99}
 
 

\bibitem{anosov}  D. V. Anosov, \emph{Geodesic flows on closed Riemannian manifolds with negative curvature},  Trudy Mat. Inst. Steklov., Vol. {\bf 90} (1967) 3 - 210

\bibitem{hedlund}G.Hedlund, \emph{The dynamics of geodesic flow}, 
Bull.Am.Math.Soc. {\bf 45} (1939) 241-246

\bibitem{hopf} E.Hopf. \emph{ Statistik der L\"osungen   
geod\"atischer Probleme vom unstabilen 
Typus.  II.}  Math.Ann. {\bf 117} (1940) 590-608



\bibitem{anosov1} D. V. Anosov and Ya. G. Sinai, \emph{Certain smooth ergodic systems}, 
Uspehi Mat. Nauk {\bf 22}
(1967), no. 5 (137), 107-172; Russian Math. Surveys {\bf22} (1967), 103-167.  


\bibitem{kolmo} A.N. Kolmogorov,  \emph{New metrical invariant of transitive dynamical 
systems and automorphisms of Lebesgue spaces}, 
Dokl. Acad. Nauk SSSR,  {\bf{119}} (1958) 861-865


\bibitem{kolmo1} A.N. Kolmogorov,  \emph{On the entropy per unit time as a metrical invariant
of automorphism}, 
Dokl. Acad. Nauk SSSR,  {\bf{124}} (1959) 754-755

\bibitem{sinai3} Ya.G. Sinai,  \emph{On the Notion of Entropy of a Dynamical System}, Doklady of Russian Academy of Sciences, {\bf{124}}  (1959)  768-771.

 
\bibitem{krilov}N.S.Krylov, \emph{Works on the foundation of statistical physics}, M.- L. Izdatelstvo
Acad.Nauk. SSSR, 1950; (Princeton University Press, 1979)
 
\bibitem{turbul}V.Arnold, \emph{Sur la g\'eom\'etrie des groupes de Lie de dimension infinie et ses 
applications en hydrodynamique des fluides parfaits}, Ann.Inst. Fourier (Grenoble) 
{\bf 16}, No 1 (1966) 319-361

\bibitem{arnoldavez}V.Arnold and A.Avez, \emph{Ergodic Problems of Classical Mechanics},
(The Mathematical physics monograph series) Benjamin (July 5, 1968), 286pp.

\bibitem{yangmillsmech} G.Savvidy, \emph{The Yang-Mills mechanics as a Kolmogorov 
K-system}, Phys.Lett.B {\bf{130}} (1983) 303

 
\bibitem{Savvidy:1982jk}
  G.~Savvidy,
\emph{Classical and Quantum Mechanics of Nonabelian Gauge Fields,}
  Nucl.\ Phys.\ B {\bf 246} (1984) 302.
   
\bibitem{body} V.Gurzadyan and G.Savvidy, \emph{Collective relaxation of stellar systems},
Astron. Astrophys. {\bf 160} (1986) 203


\bibitem{garry}G.W.Gibbons, \emph{The Jacoby-metric for timelike geodesics in 
static spacetime}, arXiv:1508.06755 [gr-qc], August 2015.


\bibitem{yer1986a}  G. Savvidy and N. Ter-Arutyunyan-Savvidy, \emph{ On the Monte Carlo simulation of physical systems}, J.Comput.Phys. {\bf 97} (1991) 566; Preprint EFI-865-16-86-YEREVAN, Jan. 1986. 13pp.
 
\bibitem{rokhlin1} V. A. Rokhlin, \emph{Metric properties of endomorphisms of compact commutative groups}, Izv. Akad. Nauk SSSR Ser. Mat.,  Volume  {\bf{28}}, Issue 4 (1964) 867- 874   

\bibitem{leonov}V. P. Leonov, \emph{On the central limit theorem for ergodic endomorphisms 
of the compact commutative groups}, Dokl. Acad. Nauk SSSR,  {\bf 124} No: 5 (1969) 980-983



\bibitem{rokhlin} V.A. Rokhlin, \emph{On the endomorphisms of compact commutative groups}, 
Izv. Akad. Nauk, vol.  {\bf{13}} (1949), p.329 

\bibitem{rokhlin2}V.A. Rokhlin, \emph{On the entropy of automorphisms of compact 
commutative groups}, 
 Teor. Ver. i Pril., vol. {\bf3}, issue 3  (1961) p. 351


\bibitem{smale}S. Smale, \emph{ Differentiable dynamical systems}. 
Bull. Am. Math. Soc. {\bf{73}} (1967) 747-817

\bibitem{sinai2}Ya. G. Sinai, \emph{Markov partitions and C-diffeomorphisms}, Funkcional. Anal, i Prilozen.
{\bf 2} (1968),64-89; Functional Anal. Appl. {\bf 2} (1968) 61-82.  

\bibitem{sinai4}Ya. G. Sinai, \emph{Proceedings of the International Congress 
of Mathematicians}, Uppsala (1963) 540-559.  


\bibitem{margulis}G. A. Margulis, \emph{Certain measures that are connected with C-flows on compact manifolds}, Funkcional. Anal, i Prilozen. {\bf 4}  (1970) 62-76; Functional Anal. Appli. {\bf 4} (1970) 55-67.

\bibitem{bowen0} R.Bowen, \emph{Equilibrium States and the Ergodic Theory of Anosov Diffeomorphisms}. (Lecture Notes in Mathematics, no. 470: A. Dold and B. Eckmann, editors). Springer-Verlag (Heidelberg, 1975), 108 pp.


\bibitem{bowen} R.Bowen, \emph{ Periodic orbits for hyperbolic flows}, Amer. J. Math.,{\bf 94} (1972), 1-30.


\bibitem{bowen1} R.Bowen, \emph{Periodic points and measures for axiom A diffeomorphisms},
Trans. Am. Math. Soc. {\bf 154}  (1971)  377-397

\bibitem{kornfeld}I. ~P. ~Kornfeld, S.~ V.~ Fomin, Y.~ G.~ Sinai,   \emph{Ergodic Theory},
Springer, 1982
 
\bibitem{gines}A.~L.~Gines, \emph{Metrical properties of the endomorphisms on m-dimensional 
torus},  Dokl. Acad. Nauk SSSR,  {\bf 138} (1961) 991-993
 
\bibitem{metropolis} N.~C.~Metropolis and S.~Ulam, 
\emph{The Monte Carlo method} , J.\ Amer. \ Statistical \ Assoc.
{\bf 44} (1949) 335-341
 
\bibitem{neuman} N.~C.~Metropolis, G.~Reitwiesner and  J.~Von Neuman, 
\emph{Statistical Treatment of Values of First 2000 Decimal Digits of 
e and of $\pi$ Calculated on the ENIAC} , Math.\ Tables and Other Aids 
to Comp. {\bf 4} (1950) 109-111 


\bibitem{neuman1} J.~Von Neuman, 
\emph{Various Techniques Used in Connection with Random Digits.
Chapter 13 of Proceedings of Symposium on "Monte Carlo Method" 
held June-July 1949 in Los Angeles.} , J.\ Res.\  Nat.\ Bur. \ Stand. \ Appl.\ Math. \ Ser.
{\bf 12} (1951) 36-38 
 
 
\bibitem{sobol}I.~M.~Sobol,  \emph{The Monte Carlo Method},  
Univ. of Chicago Press, Chicago, 1974



\bibitem{yer1986b}  N. Akopov, G. Savvidy and N. Ter-Arutyunyan-Savvidy, \emph{Matrix generator of pseudorandom numbers}, J.Comput.Phys. {\bf 97}  (1991)  573; EFI-867-18-86-YEREVAN, Jan. 1986. 8pp. 

\bibitem{mixmaxGalois}
     G. G. Athanasiu, E. G. Floratos, G. K. Savvidy
    \emph{K-system generator of pseudorandom numbers on Galois field,}
     Int. J. Mod. Phys. C {\bf 8} (1997) 555-565 .

\bibitem{pierr} P. L'Ecuyer and R. Simard, \emph{TestU01: A C Library for Empirical Testing of Random Number Generators},  ACM Transactions on Mathematical Software, {\bf 33} (2007) 1-40.

\bibitem{fred} F. James, \emph{Finally, a theory of random number generation.} "Fifth International Workshop on Mathematical Methods", CERN, Geneva, October 2001.

\bibitem{Demchik:2010fd}
  V.~Demchik,
  \emph{Pseudo-random number generators for Monte Carlo simulations on Graphics Processing Units,}
  Comput.\ Phys.\ Commun.\  {\bf 182} (2011) 692
  [arXiv:1003.1898 [hep-lat]].
 
\bibitem{falcion}M. Falcioni, L. Palatella, S. Pigolotti and A. Vulpiani, \emph{Properties making a chaotic system a good Pseudo Random Number Generator}, Phys.Rev. E {\bf 72} (2005) 016220




\bibitem{niki} N. Niki, 
     Finite field arithmetic and multidimensional uniform pseudorandom numbers (in Japanese), 
     Proc. Inst. Statist. Math. 32 (1984) 231.

 \bibitem{nied}
   H.Niederreiter,
     A pseudorandom vector generator based on finite field arithmetic, 
     Mathematica Japonica, Vol. 31, pp. 759-774, (1986)
     

\bibitem{lnbook}   R. Lidl and H. Niederreiter, Finite Fields, Addison-Wesley, Reading, MA, 1983, see also
      \emph{Finite fields, pseudorandom numbers, and quasirandom points,}
        in : Finite fields, Coding theory, and Advance in Communications and Computing. 
        (G.L.Mullen and P.J.S.Shine, eds) pp. 375-394, Marcel Dekker, N.Y. 1993.

\bibitem{konstantin} K.Savvidy, \emph{The MIXMAX random number generator},
Comput.Phys.Commun. 196 (2015) 161; \url{http://dx.doi.org/10.1016/j.cpc.2015.06.003}
  

     
\bibitem{hepforge} HEPFORGE.ORG, {http://mixmax.hepforge.org}; \\
\url{http://www.inp.demokritos.gr/~savvidy/mixmax.php}

\bibitem{cern} MIXMAX workshop:  
\url{https://indico.cern.ch/event/404547/}
 
\bibitem{root}  ROOT, Release 6.04/06 on  2015-10-13, \\
 \url{https://root.cern.ch/doc/master/mixmax_8h_source.html}
 
 
\bibitem{geant}
GEANT/CLHEP, Release 2.3.1.1, on November 10th, 2015\\
\url{http://proj-clhep.web.cern.ch/proj-clhep/}


 
\end{thebibliography}
\end{document}